\documentclass[aps,prc,reprint,groupedaddress,longbibliography,noeprint,nofootinbib,
  amsmath,amssymb,floatfix]{revtex4-2}
\usepackage[T1]{fontenc}
\usepackage{graphicx}
\usepackage{bm}
\usepackage{slashed}
\usepackage{booktabs}
\usepackage{tikz}
\usetikzlibrary{arrows.meta,decorations.markings}
\usepackage{hyperref}
\hypersetup{colorlinks=true,linkcolor=blue,citecolor=blue,urlcolor=blue}

\newcommand{\Occc}{\Omega_{ccc}}
\newcommand{\hc}{\mathrm{H.c.}}
\newcommand{\dd}{\mathrm{d}}
\newcommand{\Real}{\operatorname{Re}}
\begin{document}

\title{Nuclear bound state spectra of $\Omega$ and $\Omega_{ccc}$}
\author{Tian-Le Gao$^{1,2,4}$}\email{gaotl2025@lzu.edu.cn}
\author{Xiang Liu$^{1,2,3,4}$}\email{xiangliu@lzu.edu.cn}
\affiliation{
	$^1$School of Physical Science and Technology, Lanzhou University, Lanzhou 730000, China\\
	$^2$Lanzhou Center for Theoretical Physics, Key Laboratory of Theoretical Physics of Gansu Province, Key Laboratory of Quantum Theory and Applications of MoE and Gansu Provincial Research Center for Basic Disciplines of Quantum Physics, Lanzhou University, Lanzhou 730000, China\\
	$^3$MoE Frontiers Science Center for Rare Isotopes, Lanzhou University, Lanzhou 730000, China\\
	$^4$Research Center for Hadron and CSR Physics, Lanzhou University and Institute of Modern Physics of CAS, Lanzhou 730000, China
}

\begin{abstract}
We investigate the nuclear bound-state spectra of $\Omega$ and $\Omega_{ccc}$
in the unquenched description, using in-medium hadron masses from the
quark-meson coupling model. The hadronic-loop-induced scalar potentials are
constructed in the local density approximation and combined with the Coulomb
field to calculate energies and radial wave functions for six nuclei from
${}^{12}\mathrm{C}$ to ${}^{208}\mathrm{Pb}$. For $\Omega$, the scalar and
Coulomb attractions yield a hydrogen-like grouping of shallow excited states
in light nuclei at weak scalar coupling; stronger attraction or a larger
nuclear charge promotes interior localization and approximately
harmonic-oscillator-like level patterns. For $\Omega_{ccc}$, binding requires
scalar attraction to overcome Coulomb repulsion. For sufficiently strong
attraction, the spectra in light nuclei show oscillator-like grouping,
whereas an off-center potential minimum in heavy nuclei compresses the $1s$,
$1p$, and $1d$ levels relative to the $2s$ radial excitation. A cosh scalar
potential with a shared nuclear geometry, combined with a uniform-sphere
Coulomb field matched to the root-mean-square radius of the assumed charge
profile, reproduces the bound-state energies and excitation gaps with
few-MeV and sub-MeV root-mean-square deviations, respectively. A scan of the
scalar depth indicates that, within the present approximations, relatively
light nuclei give the deepest $\Omega_{ccc}$ binding.
\end{abstract}
\maketitle
\raggedbottom

\section{Introduction}
\label{sec:introduction}
Studies of low-energy hadron interactions provide a way to investigate the
nonperturbative problem of strong interaction. A nucleus provides a setting in
which to study how these interactions modify the properties of an embedded
hadron~\cite{Saito2007}. The possibility of binding heavy quarkonia to nuclei
was investigated in early studies of gluon-mediated
forces~\cite{Brodsky1990NuclearQuarkonium,Luke1992QuarkoniumNuclei}. Lattice
QCD calculations have also explored the interactions of strange and charm
quarkonia with light nuclei at unphysical light-quark
masses~\cite{Beane2015QuarkoniumNuclei}. When bound states form, their binding
energies and level spacings probe the strength and radial dependence of the
effective hadron--nucleus interaction.

To relate the hadron--nucleus interaction to changes in the structure of the
embedded hadron, we use the quark--meson coupling (QMC) model. The nuclear
scalar and vector meson fields act on confined light quarks and antiquarks.
The scalar field changes their effective masses and hence the hadron bag
energy. The response of the nucleons to the fields enters the self-consistent
determination of the nuclear mean
fields~\cite{Guichon1988,Guichon1996,Saito1997FiniteMasses,Saito2007}. With
direct couplings restricted to $u$ and $d$ flavors in this
prescription~\cite{Tsushima2003,Tsushima2019}, all-strange and all-heavy
hadrons require an indirect mechanism for a mass shift. In an unquenched
description, quark--antiquark pair creation couples the valence state to
hadronic channels. Virtual propagation in these channels contributes to the
self-energy even below the threshold for producing the intermediate hadrons.
The importance of this coupling near thresholds is illustrated by the
$D\bar{K}$ channel for the $D^*_{s0}(2317)$~\cite{VanBeveren2003ScalarCharm}
and by open-flavor channels in
quarkonium~\cite{Ono1984Continuum,Kalashnikova2005Coupled,Li2009Coupled,Danilkin2010Pole}
(see the review article~\cite{Bai2027Unquenched} and references therein for
more details). In nuclear matter, intermediate hadrons with light constituents
acquire density-dependent masses through the QMC scalar field. The resulting
change in the loop self-energy shifts the mass of the external hadron.

This mechanism has been applied to the predominantly $\phi$ meson and to heavy
quarkonia. The response of kaons and antikaons to nuclear matter has been
studied in QMC~\cite{Tsushima1998Kaons}. Calculations with in-medium $K\bar K$
loops relate changes in the $\phi$ mass and decay width to the possible
formation of $\phi$--nucleus bound
states~\cite{CobosMartinez2017PhiMatter,CobosMartinez2017PhiNuclei,Mondal2025PhiNuclei}.
Related studies of $J/\psi$, $\eta_c$, and bottomonium use in-medium
open-flavor meson
loops~\cite{Krein2011Loops,Tsushima2011Bound,Cobos2020Etac,Cobos2022Bottomonia}.
For the $\chi_{cJ}(1P)$ triplet (with total angular momentum $J=0,1,2$),
QMC-based calculations of the density-dependent self-energies~\cite{Zhang2026}
have been extended to finite nuclei, where a cosh potential with a common
nuclear geometry gives a compact description of the calculated
spectra~\cite{Gao2026}.

The application of this approach to finite-nucleus spectra of all-strange and triply charmed baryons has received less attention. Virtual baryon--meson channels have been included in studies of the vacuum $\Omega$ mass~\cite{GarciaTecocoatzi2017BaryonLoops,Ren2014Decuplet}. Lattice QCD calculations have found attraction in the spin-2 $N\Omega$ $S$ wave and predicted a shallow quasibound state near the physical point~\cite{Etminan2014NOmega,Iritani2019NOmega}. Proton--$\Omega$ correlations measured by ALICE also support an attractive interaction~\cite{ALICE2020HadronInteraction}. The binding of $\Omega$ in light nuclear systems has been investigated using lattice-based interactions~\cite{Garcilazo2019OmegaNN,Etminan2020OmegaAlpha,Filikhin2026OmegaLightNuclei}.

For $\Occc$, lattice studies of the vacuum spectrum~\cite{Dhindsa2025} are now complemented by calculations of its interaction with nucleons. At physical quark masses, HAL QCD finds overall attraction in the spin-1 and spin-2 $N\Occc$ $S$ waves, with neither channel supporting a two-body bound state~\cite{Zhang2025NOmegaccc}. Few-body calculations have since examined $\Occc np$ and $\alpha\alpha\Occc$ systems~\cite{Filikhin2025OmegacccNN,ZhouLiu2026}. These studies motivate examining whether $\Occc$ can bind in finite nuclei in the presence of Coulomb repulsion and how the spectrum varies with the nuclear core. The ground-state baryons $\Omega$ and $\Occc$ both have $J^P=3/2^+$. Their charges, $-e$ and $+2e$, allow a comparison of spectra in which the Coulomb interaction reinforces or opposes the scalar attraction, respectively.

In this work, we calculate the in-medium masses of $\Omega$ and $\Occc$ from the $\Xi^{(*)}\bar K^{(*)}$ and $\Xi_{cc}^{(*)}D^{(*)}$ loops, respectively, with intermediate hadron masses from QMC. We compare the resulting baryon masses with the lowest baryon--meson thresholds in cold symmetric nuclear matter. Within the local density approximation, we use the mass shifts to construct scalar potentials for six nuclei from ${}^{12}\mathrm{C}$ to ${}^{208}\mathrm{Pb}$ and include the corresponding Coulomb fields. The resulting bound state energies and radial wave functions are used to examine how scalar attraction and the Coulomb interaction affect the level spacings and localization. We also test a cosh scalar potential combined with a Coulomb term matched to the root-mean-square radius of the assumed nuclear charge profile. By varying the scalar depth, we explore which nuclei favor deep $\Occc$ binding within this parametrization.

This paper is organized as follows:  Section~\ref{sec:framework} describes the hadronic loops, their QMC input, and the finite-nucleus bound state equation. Section~\ref{sec:results} presents the in-medium masses, potentials, spectra, and radial wave functions. Section~\ref{sec:cosh} tests the cosh parametrization and examines the dependence of $\Occc$ binding on the nuclear core and scalar depth. Section~\ref{sec:summary} summarizes the results, and Appendix~\ref{app:spectra} tabulates the reference and cosh spectra.

\section{Theoretical framework}
\label{sec:framework}

We describe the nuclear medium with the quark--meson coupling (QMC) model~\cite{Guichon1988,Guichon1996,Saito2007}, using the finite nucleus prescription of Ref.~\cite{Gao2026}. In this framework, the nuclear mean fields couple to light valence quarks, whereas their direct couplings to strange and charm quarks are neglected. The $J^P=3/2^+$ baryons $\Omega$ and $\Occc$ therefore acquire strong interaction mass shifts through virtual baryon--meson channels containing light constituents. These shifts provide the strong potentials for the subsequent nuclear bound state calculation.

For a light quark or antiquark in a hadron bag, the equation in uniform, isospin-symmetric matter is
\begin{equation}
 \left[i\slashed\partial-m_q^*
       \mp\gamma^0g_\omega^q\omega\right]
 \begin{pmatrix}\psi_q\\\psi_{\bar q}\end{pmatrix}=0,
 \label{eq:quarks}
\end{equation}
where $m_q^*=m_q-g_\sigma^q\sigma$ for $q=u,d$, and the upper (lower) sign refers to the quark (antiquark). Here $\sigma$ and $\omega$ are the scalar and temporal isoscalar vector fields; $g_\sigma^q$ and $g_\omega^q$ are their quark couplings. The other constituent masses remain $m_s^*=m_s$ and $m_c^*=m_c$. The effective mass of an intermediate hadron $h$ follows from the stationary MIT bag energy~\cite{Tsushima2003,Tsushima2019},
\begin{equation}
 \begin{aligned}
 &m_h^*(\sigma)=
 \frac{\sum_f n_{fh}\omega_{fh}^*-z_h}{R_h^*}
       +\frac{4\pi}{3}\mathcal B(R_h^*)^3,\\
 &\left.\frac{\partial m_h^*(\sigma,R)}{\partial R}
 \right|_{R=R_h^*}=0.
 \end{aligned}
 \label{eq:bagmass}
\end{equation}
The constituent energies are $\omega_{fh}^*=\sqrt{x_{fh}^{*2}+(m_f^*R_h^*)^2}$, with $x_{fh}^*$ fixed by the usual bag boundary condition. The quantities $n_{fh}$, $\mathcal B$, and $z_h$ denote the constituent multiplicity, bag constant, and phenomenological bag-energy correction, respectively; the bag parameters are calibrated in vacuum.

For each spherical nuclear core, the occupied nucleon states and the $\sigma$, $\omega$, $\rho$, and Coulomb fields are solved self-consistently~\cite{Guichon1996,Gao2026}. The QMC-I scalar response is specified by
\begin{equation}
 \begin{aligned}
 M_N^*(\sigma)&=M_N-g_\sigma\sigma
                     +\frac{a_N}{2}(g_\sigma\sigma)^2,\\
 (-\nabla^2+m_\sigma^2)\sigma
     &=g_\sigma(1-a_Ng_\sigma\sigma)\rho_s,
 \end{aligned}
 \label{eq:core}
\end{equation}
where $g_\sigma$ is the zero field nucleon coupling, $a_N$ parametrizes its scalar response, and $\rho_s$ is the nucleon scalar density. The resulting local field $\sigma_A(r)$ determines the intermediate masses through $m_h^{*,A}(r)=m_h^*[\sigma_A(r)]$. The core is held fixed when the additional baryon is introduced.

The $\Omega$ self energy includes the channels $\Xi\bar K$, $\Xi^*\bar K$, $\Xi\bar K^*$, and $\Xi^*\bar K^*$. Keeping only the lowest order interaction structures~\cite{AlievP2010,Bertilsson2024,AlievODV2010,AlievDDV2010}, we write
\begin{subequations}\label{eq:interactions}
\begin{align}
 \mathcal L_{\Xi\bar K\Omega}
   &=g_P\bar\Xi(\partial_\mu K)\Omega^\mu+\hc,
 \label{eq:lagpk}\\
 \mathcal L_{\Xi^*\bar K\Omega}
   &=-f_P\bar\Xi^{*\mu}\gamma^\nu\gamma_5
       (\partial_\nu K)\Omega_\mu+\hc,
 \label{eq:lagpkstarb}\\
 \mathcal L_{\Xi\bar K^*\Omega}
   &=g_V\bar\Xi\gamma^\nu\gamma_5\Omega^\mu
        K_{\mu\nu}^*+\hc,
 \label{eq:lagvk}\\
 \mathcal L_{\Xi^*\bar K^*\Omega}
   &=f_V\bar\Xi^{*\mu}\gamma^\nu\Omega_\mu
        K_\nu^*+\hc,
 \label{eq:lagvkstarb}
\end{align}
\end{subequations}
where $K_{\mu\nu}^*=\partial_\mu K_\nu^* -\partial_\nu K_\mu^*$, $K^{(*)}=(\bar K^{(*)})^\dagger$, and isospin indices are suppressed. The coupling coefficients are taken to be independent of the nuclear medium.

\begin{figure}[t]
 \centering
 \begin{tikzpicture}[
     loop line/.style={line width=1.05pt},
     fermion/.style={
       loop line,
       postaction={decorate},
       decoration={markings,mark=at position 0.5 with
         {\arrow{Stealth[length=4.3pt,width=3.6pt]}}}
     },
     vertex/.style={circle,fill=black,inner sep=1.4pt},
     particle/.style={font=\normalsize,inner sep=1pt},
     panel/.style={font=\small,anchor=west,inner sep=0pt}
   ]
  \def\R{0.60}
  \foreach \x/\panelid/\parent/\baryon/\meson in
    {0/a/{\Omega}/{\Xi^{(*)}}/{K^{(*)}},
     3.85/b/{\Occc}/{\Xi_{cc}^{(*)}}/{D^{(*)}}} {
   \begin{scope}[shift={(\x,0)}]
    \draw[fermion] (-1.70,0) -- (-\R,0);
    \draw[fermion] (-\R,0) arc[start angle=180,end angle=0,radius=\R];
    \draw[fermion] (\R,0) -- (1.70,0);
    \draw[loop line] (-\R,0) arc[start angle=180,end angle=360,radius=\R];
    \node[vertex] at (-\R,0) {};
    \node[vertex] at (\R,0) {};
    \node[panel] at (-1.70,0.86) {(\panelid)};
    \node[particle] at (-1.17,0.32) {$\parent$};
    \node[particle] at (1.17,0.32) {$\parent$};
    \node[particle,anchor=south] at (0,0.65) {$\baryon$};
    \node[particle,anchor=north] at (0,-0.65) {$\meson$};
   \end{scope}
  }
 \end{tikzpicture}
 \caption{Virtual baryon--meson contributions to the in-medium
 self-energies of (a) $\Omega$ and (b) $\Occc$. Arrows on the baryon
 lines indicate baryon-number flow. The
 $K^{(*)}$ field label in (a) corresponds to an intermediate
 $\bar K^{(*)}$ state. Each $(*)$ independently denotes the unstarred
 or starred state. The QMC scalar field modifies the intermediate
 hadron masses.}
 \label{fig:hadronic-loops}
\end{figure}
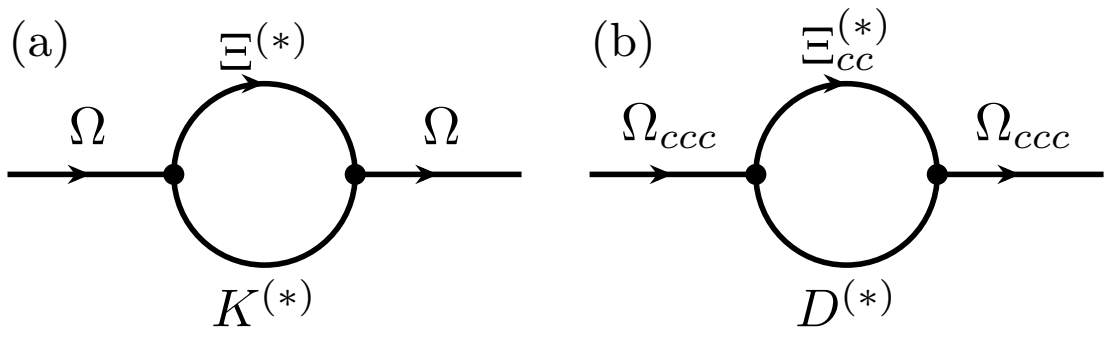

For $\Occc$, we assume exact $SU(4)$ relations among the corresponding couplings. Its four channels and vertices follow by replacing $\Omega\to\Occc$, $\Xi^{(*)}\to\Xi_{cc}^{(*)}$, and $\bar K^{(*)}\to D^{(*)}$, with the conjugate fields replaced accordingly in Eq.~\eqref{eq:interactions}. The coupling coefficients are unchanged, while the strange and charmed masses remain distinct in the propagators. Figure~\ref{fig:hadronic-loops} illustrates the virtual channels for both baryons.

Let $H=\Omega,\Occc$ and $p=(W,\bm0)$, with $W=\sqrt{p^2}>0$. The positive-energy, spin-$3/2$ projected self energy is
\begin{align}
 \Sigma_H^+(W;\sigma,\alpha)
 ={}&-i\sum_{C\in\mathcal C_H}2
       \int\frac{\dd^4k}{(2\pi)^4}
       \nonumber\\
 &\times\frac{\mathcal N_C(p,k)\,
       \mathcal F_B(k^2;\sigma)\mathcal F_M(q^2;\sigma)}
 {(k^2-m_B^{*2})(q^2-m_M^{*2})},
 \label{eq:selfenergy}
\end{align}
Here $B$ and $M$ label the intermediate baryon and meson, $q=p-k$, and $\mathcal C_H$ is the four channel set. The factor 2 accounts for the two charge channels with single charge channel couplings. Let $\mathcal T_{C,\nu\mu}(p,k)$ be the unprojected channel numerator containing the vertices and signed propagator numerators, after extracting the common factors displayed in Eq.~\eqref{eq:selfenergy}. Its projection is
\begin{equation}
 \mathcal N_C(p,k)=\frac{1}{4}\operatorname{Tr}_D\!
 \left[\Lambda_+(p)\mathcal P_{3/2}^{\mu\nu}(p)
       \mathcal T_{C,\nu\mu}(p,k)\right],
 \label{eq:projected-numerator}
\end{equation}
where $\operatorname{Tr}_D$ is the Dirac trace and $1/4$ averages over the four physical spin states. The spin-$3/2$ and positive energy projectors, in the metric $g^{\mu\nu}=\operatorname{diag}(1,-1,-1,-1)$, are
\begin{equation}
 \begin{aligned}
 \mathcal P_{3/2}^{\mu\nu}(p)
 &=g^{\mu\nu}-\frac{1}{3}\gamma^\mu\gamma^\nu
   -\frac{\slashed{p}\gamma^\mu p^\nu
          +p^\mu\gamma^\nu\slashed{p}}{3p^2},\\
 \Lambda_+(p)&=\frac{\slashed{p}+W}{2W}
               =\frac{1+\gamma^0}{2}.
 \end{aligned}
 \label{eq:spin-energy-projectors}
\end{equation}
The last equality holds in the rest frame. For intermediate spin-$3/2$ baryons we retain the full Rarita--Schwinger numerator~\cite{Pascalutsa2000}. Only scalar mass modifications enter the loops: the opposite light quark and antiquark vector shifts cancel in the homogeneous background when expressed in kinetic momenta.

Each intermediate line is regulated by a dipole factor,
\begin{equation}
 \begin{aligned}
 \mathcal F_h(s;\sigma)
 &=\left[\frac{m_h^{*2}(\sigma)-\Lambda_h^2(\sigma)}
                  {s-\Lambda_h^2(\sigma)}\right]^2,\\
 \Lambda_h(\sigma)&=m_h^*(\sigma)+\alpha\Lambda_{\rm QCD},
 \end{aligned}
 \label{eq:regulator}
\end{equation}
with $\Lambda_{\rm QCD}=220$ MeV and dimensionless cutoff parameter $\alpha$~\cite{Zhang2026}. The same prescription is used in vacuum and in the medium. Defining $\Delta M_H=M_H^*-M_H$, the vacuum-subtracted pole equation is
\begin{align}
 \Delta M_H(\sigma;\alpha)
 ={}&\Real\Sigma_H^+(M_H+\Delta M_H;\sigma,\alpha)
       \nonumber\\
    &-\Real\Sigma_H^+(M_H;0,\alpha),
 \label{eq:subtractedmass}
\end{align}
where $M_H$ is the vacuum pole mass. We select the solution continuously connected to $\Delta M_H=0$ at $\sigma=0$.

Within the local approximation, the central potential in a nucleus of mass number $A$ and charge $Z$ is
\begin{equation}
 U_H^A(r;\alpha)=\Delta M_H[\sigma_A(r);\alpha]
                      +Q_He\Phi_A(r),
 \label{eq:totalpotential}
\end{equation}
where $Q_\Omega=-1$ and $Q_{\Occc}=+2$. Here $\Phi_A(r)$ is the electrostatic potential generated by the nuclear proton density $\rho_p^A$. It obeys $-\nabla^2\Phi_A=e\rho_p^A$ and satisfies $e\Phi_A(r)\to Z\alpha_{\rm em}/r$ asymptotically, with $\alpha_{\rm em}=e^2/(4\pi)$. The Coulomb contribution is attractive for $\Omega$ and repulsive for $\Occc$. The radial bound state equation is
\begin{equation}
 \left[-\frac{\mathcal{D}_\ell}{2\mu_{H,A}}+U_H^A(r;\alpha)\right]
 u_{n\ell}(r)=E_{n\ell}^{H,A}(\alpha)u_{n\ell}(r),
 \label{eq:schrodinger}
\end{equation}
where $\mathcal{D}_\ell=\mathrm{d}^2/\mathrm{d}r^2-\ell(\ell+1)/r^2$
is the radial differential operator acting on the reduced radial wave
function $u_{n\ell}(r)$. The reduced mass is
$\mu_{H,A}=M_HM_A/(M_H+M_A)$, with $M_A$ the nuclear core mass. The reduced radial wave function vanishes at the origin and decays at infinity. The eigenvalue is measured from the free baryon--core threshold, so that a bound state has $E_{n\ell}<0$ and separation energy $B_{n\ell}=-E_{n\ell}$. This central, nonrelativistic treatment neglects nuclear polarization, spin-dependent splittings, and absorptive widths.

\section{Numerical results}
\label{sec:results}
We use $M_\Omega=1672$ MeV from the Particle Data Group (PDG)~\cite{PDG2024} and $M_{\Occc}=4793$ MeV from the quark-model calculation of Ref.~\cite{Zhou2025}. The intermediate strange baryon and meson masses are based on PDG values. A common representative mass is adopted for the members of each isospin multiplet. The vacuum mass inputs are collected in Table~\ref{tab:vacuum-masses}.

For the intermediate doubly charmed baryons, the representative value $m_{\Xi_{cc}}=3621$ MeV is based on the mean of the measured masses of the two charge states~\cite{LHCb2020Xicc,LHCb2026Xicc}. The spin-partner input $m_{\Xi_{cc}^*}=3704$ MeV is estimated by adding the central lattice-QCD hyperfine splitting from Brown et al.~\cite{Brown2014} to the measured $\Xi_{cc}^{++}$ mass.

\begin{table}[t]
 \caption{Vacuum masses in MeV, rounded to the nearest integer for display. The sources and the estimate of the $\Xi_{cc}^*$ mass are specified in the text.}
 \label{tab:vacuum-masses}
 \centering
 \begin{tabular*}{\columnwidth}{@{\extracolsep{\fill}}cccc}
  \toprule\toprule
  Hadron & Mass & Hadron & Mass \\
  \midrule
  $\Omega$ & 1672 & $\Occc$ & 4793 \\
  $\Xi$ & 1318 & $\Xi_{cc}$ & 3621 \\
  $\Xi^*$ & 1532 & $\Xi_{cc}^*$ & 3704 \\
  $\bar K$ & 496 & $D$ & 1867 \\
  $\bar K^*$ & 894 & $D^*$ & 2009 \\
  \bottomrule\bottomrule
 \end{tabular*}
\end{table}

The nuclear backgrounds use the QMC-I parametrization employed in earlier finite-nucleus studies~\cite{Guichon1996,Saito2007,Gao2026}, with $a_N=8.8\times10^{-4}$ MeV$^{-1}$, $(g_\sigma^2,g_\omega^2,g_\rho^2)/(4\pi)=(3.12,5.31,6.93)$, and $(m_\sigma,m_\omega,m_\rho)=(418,783,770)$ MeV. The bag calculation uses $m_{u,d}=5$ MeV, $m_s=250$ MeV, $m_c=1300$ MeV, and $\mathcal B^{1/4}=170$ MeV; the light quark scalar coupling in the finite-nucleus convention is $g_\sigma^q=4.32$. These parameters are held fixed throughout the calculation.

In the vertex convention of Eq.~\eqref{eq:interactions}, the single-charge-channel couplings are $g_P=8.0~\mathrm{GeV}^{-1}$, $f_P=8.78~\mathrm{GeV}^{-1}$, $g_V=-14.0~\mathrm{GeV}^{-1}$, and $f_V=-23.5$. Here $g_P$, $g_V$, and $f_V$ use the central values of the corresponding light cone QCD sum rule results~\cite{AlievP2010,AlievODV2010,AlievDDV2010}, retaining only the structures written in Eq.~\eqref{eq:interactions}. For the decuplet--decuplet pseudoscalar vertex, we set $f_P=H_A/(\sqrt{6}f_0)$ with the benchmark choice $H_A=2$ and $f_0=93$ MeV~\cite{Bertilsson2024}. The same coefficients are assigned to the charmed vertices under the assumed $SU(4)$ relations. With these inputs fixed, the cutoff parameter $\alpha$ is the only free parameter in the present calculation. We consider the representative values $\alpha=1,1.5,2,2.5,3$.

Table~\ref{tab:mass-shifts} gives the resulting mass shifts in symmetric nuclear matter at $\rho_0=0.15$ fm$^{-3}$. Both baryons undergo a downward mass shift whose magnitude increases with $\alpha$. Over the chosen interval, $-\Delta M_\Omega$ grows from 8.39 to 59.16 MeV, while $-\Delta M_{\Occc}$ grows from 6.44 to 65.02 MeV. These values characterize the scalar attraction generated by the loops.

\begin{table}[t]
 \caption{Mass shifts $\Delta M_H=M_H^*-M_H$ at normal
nuclear matter density $\rho_0=0.15$ fm$^{-3}$, in MeV.}
 \label{tab:mass-shifts}
 \centering
 \begin{tabular*}{\columnwidth}{@{\extracolsep{\fill}}cccccc}
  \toprule\toprule
  $H\backslash\alpha$ & 1 & 1.5 & 2 & 2.5 & 3 \\
  \midrule
  $\Omega$ & $-8.39$ & $-18.60$ & $-32.47$ & $-46.70$ & $-59.16$ \\
  $\Occc$ & $-6.44$ & $-20.11$ & $-37.26$ & $-52.86$ & $-65.02$ \\
  \bottomrule\bottomrule
 \end{tabular*}
\end{table}

\begin{figure}[t]
 \centering
 \includegraphics[width=\columnwidth]{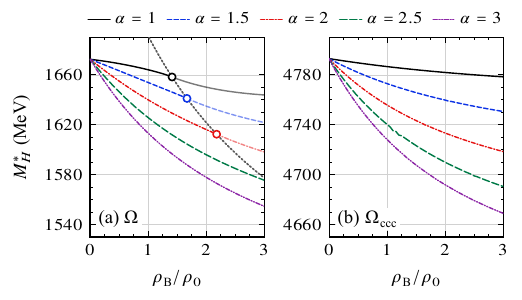}
 \caption{In-medium masses $M_H^*$ of (a) $\Omega$ and (b) $\Occc$
 in symmetric nuclear matter, with $\rho_0=0.15~\mathrm{fm}^{-3}$.
 The shared legend gives the five cutoff values.
 The gray dotted line in (a) shows the $\Xi\bar K$ threshold
 $m_\Xi^*+m_{\bar K}^*$ within the plotted mass range.
 Open circles mark its intersections with the $\Omega$ mass curves.
 In (b), the $\Occc$ mass remains below the $\Xi_{cc}D$ threshold
 for all five cutoffs throughout the plotted density range.}
 \label{fig:nuclear-matter-masses}
\end{figure}

Figure~\ref{fig:nuclear-matter-masses} shows that the effective masses of both baryons generally decrease with increasing density. At a given density, the mass reduction grows with $\alpha$. In Fig.~\ref{fig:nuclear-matter-masses}(a), the $\Omega$ mass curves cross the $\Xi\bar K$ threshold at $\rho_B/\rho_0\simeq1.41$, $1.67$, and $2.18$ for $\alpha=1$, $1.5$, and $2$, respectively. For each of these cutoffs, the decay $\Omega\to\Xi\bar K$ is kinematically allowed in uniform matter above the corresponding crossing density. The channel remains closed for $\alpha=2.5$ and $3$ up to $3\rho_0$. Over the same density range, the $\Occc$ mass remains well below the $\Xi_{cc}D$ threshold for all five cutoffs. The six finite cores reach at most about $1.33\rho_0$, and the $\Xi\bar K$ channel remains closed when evaluated with the local masses in these nuclei. Its opening at higher density therefore does not affect the bound state calculations below.

\begin{figure}[t]
 \centering
 \includegraphics[width=\columnwidth]{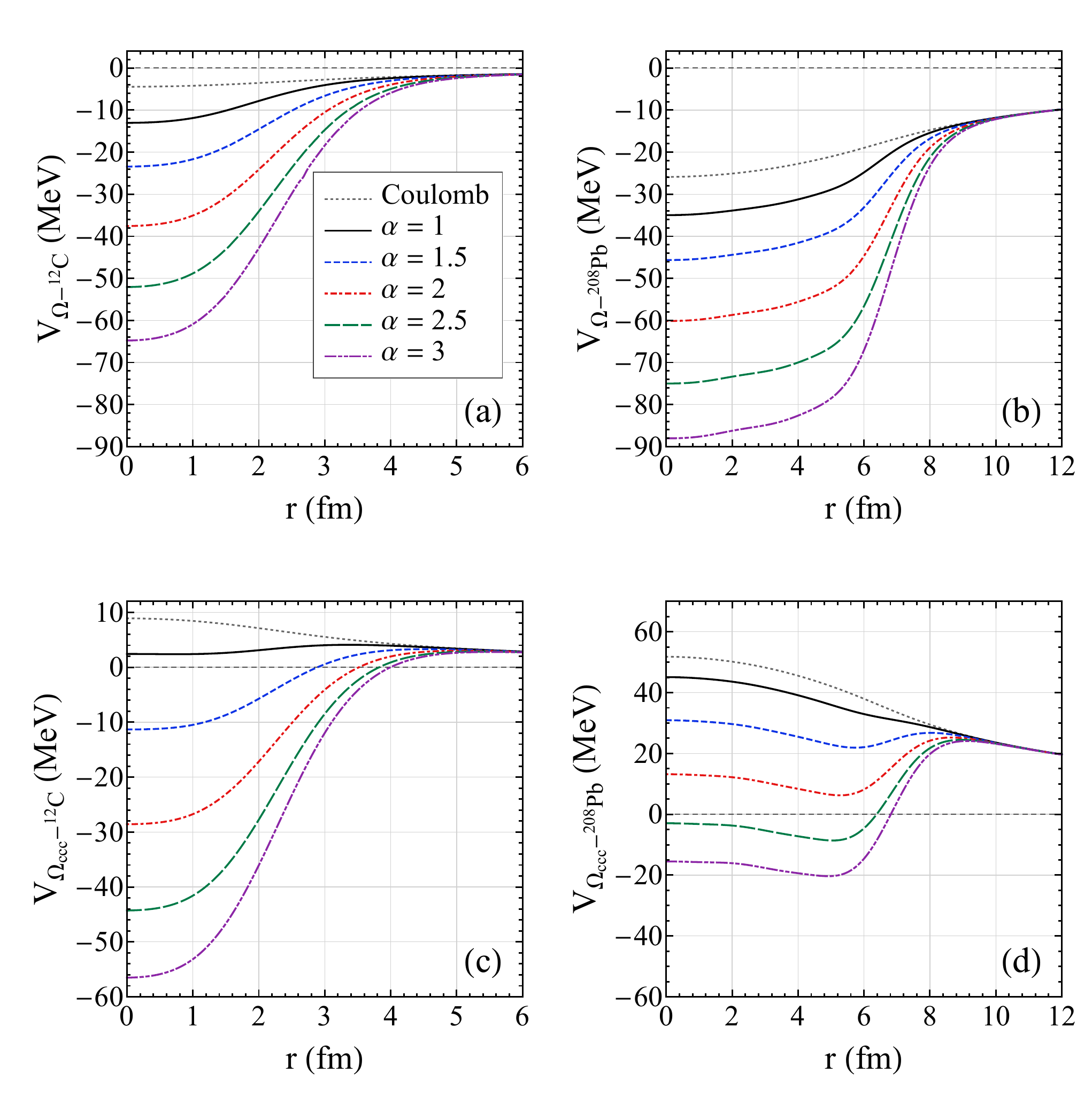}
 \caption{Total potentials $U_H^A(r;\alpha)$ of
Eq.~\eqref{eq:totalpotential}, denoted by $V_{H-A}$ on the axes, for (a) $\Omega$--${}^{12}\mathrm{C}$, (b) $\Omega$--${}^{208}\mathrm{Pb}$, (c) $\Occc$--${}^{12}\mathrm{C}$, and (d) $\Occc$--${}^{208}\mathrm{Pb}$. The colored curves correspond to the five values of $\alpha$ indicated in the shared legend in (a). Panels (c) and (d) use different vertical scales. The gray dotted curves show the Coulomb contribution alone, with charges $Q_\Omega=-1$ and $Q_{\Occc}=+2$. The curves are smoothed for display; the eigenvalue calculations use the original tabulated potentials.}
 \label{fig:potentials}
\end{figure}

For finite nuclei, we consider ${}^{12}\mathrm{C}$, ${}^{16}\mathrm{O}$, ${}^{40}\mathrm{Ca}$, ${}^{90}\mathrm{Zr}$, ${}^{197}\mathrm{Au}$, and ${}^{208}\mathrm{Pb}$. Figure~\ref{fig:potentials} shows the total potentials for ${}^{12}\mathrm{C}$ and ${}^{208}\mathrm{Pb}$ as representative light and heavy nuclei. Increasing $\alpha$ strengthens the scalar attraction, whereas the Coulomb field of a given core is unchanged. For $\Omega$, the two contributions are both attractive: the scalar term deepens the interior well and the Coulomb term supplies a long range attractive tail. For $\Occc$, the repulsive Coulomb term opposes the scalar attraction. The competition is particularly strong in ${}^{208}\mathrm{Pb}$, where a sufficiently deep scalar potential is needed to produce a negative well. Since Coulomb repulsion is strongest near the center, the minimum of the total potential can occur at a nonzero radius.

\begin{figure*}[tp]
  \centering
  \includegraphics[width=\textwidth]{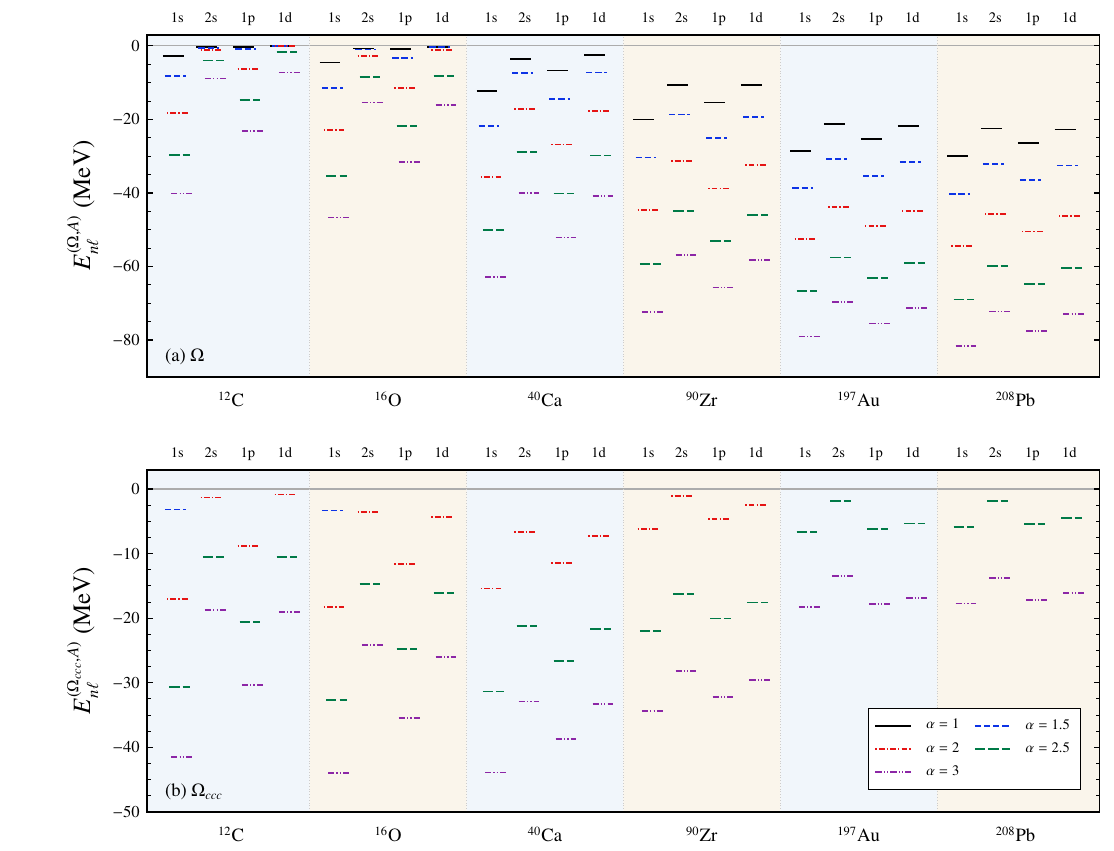}
  \caption{Selected bound state energies $E_{n\ell}^{(H,A)}$
for (a) $\Omega$ and (b) $\Occc$ in ${}^{12}\mathrm{C}$, ${}^{16}\mathrm{O}$, ${}^{40}\mathrm{Ca}$, ${}^{90}\mathrm{Zr}$, ${}^{197}\mathrm{Au}$, and ${}^{208}\mathrm{Pb}$. The five colors and line styles denote $\alpha=1$, $1.5$, $2$, $2.5$, and $3$. Alternating background tints distinguish neighboring nuclear groups. The orbital labels specify the columns within each group; small horizontal offsets separate nearly coincident levels. Energies are measured relative to the free baryon--core threshold, indicated by the horizontal line at $E=0$. Blank entries denote the absence of a negative-energy solution for the requested orbital; in particular, no displayed $\Occc$ state is bound at $\alpha=1$. The Coulomb terms use charges $Q_\Omega=-1$ and $Q_{\Occc}=+2$.}
  \label{fig:omega-omegaccc-level-spectra}
\end{figure*}

The selected bound state energies obtained from Eq.~\eqref{eq:schrodinger} are displayed in Fig.~\ref{fig:omega-omegaccc-level-spectra}. The corresponding numerical energies for both baryons are tabulated in Tables~\ref{tab:app-omega-charged} and~\ref{tab:app-ccc-charged} of Appendix~\ref{app:spectra}. We label the nuclear orbitals by $n\ell$, where $n=1,2,\ldots$ is the radial quantum number (corresponding to $n-1$ radial nodes) and $\ell$ is the orbital angular momentum quantum number. Spin-dependent splittings are not resolved. All four displayed $\Omega$ levels are bound for every core and cutoff considered. The $\Occc$ spectrum is more selective: no negative energy state is obtained at $\alpha=1$; at $\alpha=1.5$, only the $1s$ levels in ${}^{12}\mathrm{C}$ and ${}^{16}\mathrm{O}$ are bound. At $\alpha=2$, binding occurs in the four cores from ${}^{12}\mathrm{C}$ to ${}^{90}\mathrm{Zr}$, while the displayed states in ${}^{197}\mathrm{Au}$ and ${}^{208}\mathrm{Pb}$ remain unbound. All four selected orbitals are bound in all six cores at $\alpha=2.5$ and 3.

\begin{figure*}[tp]
 \centering
 \includegraphics[width=\textwidth]{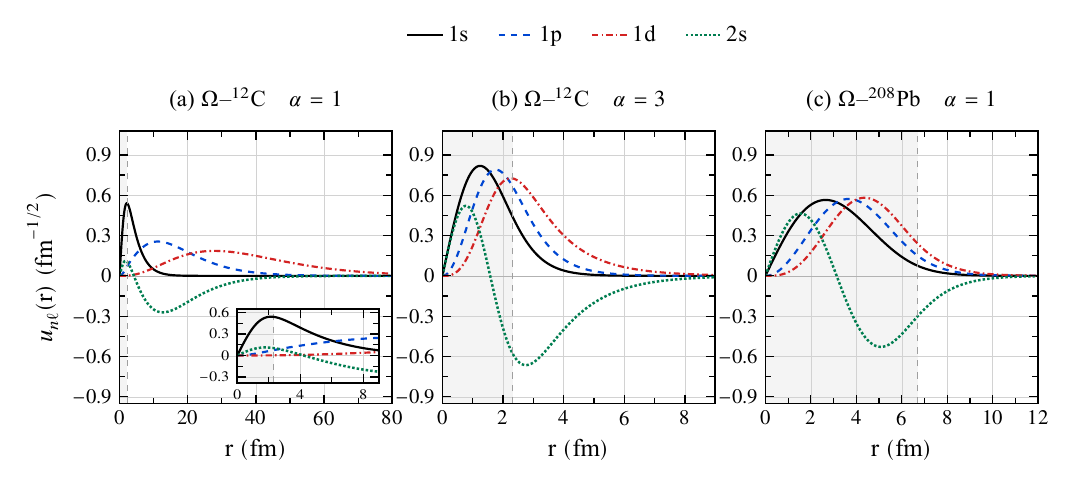}
 \caption{Reduced radial wave functions $u_{n\ell}(r)$ for
(a) $\Omega$--${}^{12}\mathrm{C}$ at $\alpha=1$, (b) $\Omega$--${}^{12}\mathrm{C}$ at $\alpha=3$, and (c) $\Omega$--${}^{208}\mathrm{Pb}$ at $\alpha=1$. Each function is normalized by $\int_0^\infty |u_{n\ell}(r)|^2\dd r=1$. The inset in (a) resolves the small-$r$ region. The shaded region and gray dashed line mark $r\leq R_{1/2}$ and $r=R_{1/2}$, respectively, where the baryon density falls to half its maximum on the outer side of the core. This radius is a reference for localization. Panels (a) and (b) compare two cutoffs for the same nuclear core, and panels (a) and (c) compare two cores at the same cutoff.}
 \label{fig:omega-waves}
\end{figure*}

We first discuss the $\Omega$ spectrum. In a light nucleus at small $\alpha$, the extended excited states mainly probe the exterior Coulomb field. The point Coulomb limit has $E_{n\ell}\propto-(n+\ell)^{-2}$ in the present notation, so that $2s$ and $1p$ belong to the same principal shell. For $\Omega$--${}^{12}\mathrm{C}$ at $\alpha=1$, their energies are $-0.43$ and $-0.37$ MeV, respectively, whereas the $1d$ level lies at $-0.15$ MeV. The corresponding wave functions in Fig.~\ref{fig:omega-waves}(a) extend far beyond the nuclear density distribution: their rms radii are 16.43, 15.69, and 34.76 fm for $2s$, $1p$, and $1d$. The $1s$ state is substantially more localized, with an rms radius of 3.49 fm, and is shifted to $-2.72$ MeV by its overlap with the attractive interior. The hydrogen-like description therefore applies primarily to the shallow excited states, rather than to the entire spectrum.

\begin{figure}[tp]
 \centering
 \includegraphics[width=\columnwidth]{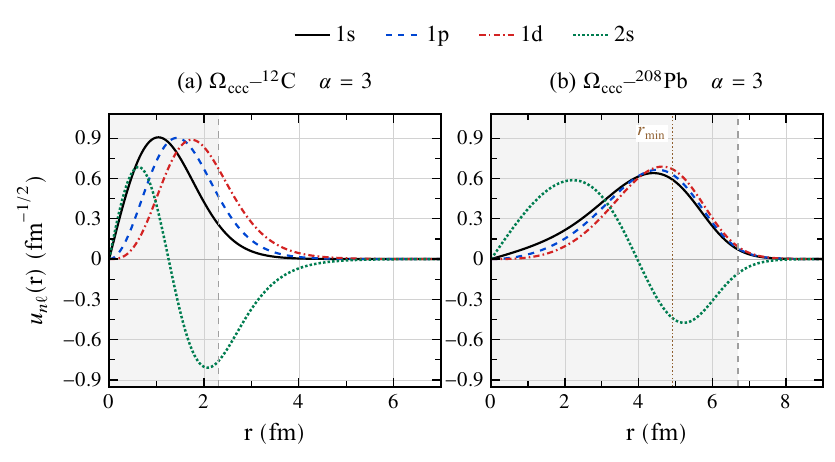}
 \caption{Normalized reduced radial wave functions for
(a) $\Occc$--${}^{12}\mathrm{C}$ and (b) $\Occc$--${}^{208}\mathrm{Pb}$ at $\alpha=3$. The normalization, curves, and half-density shading follow Fig.~\ref{fig:omega-waves}. The brown dotted line in (b) locates the minimum of the total central potential, $r_{\min}\simeq4.92$ fm. It lies inside the half-density radius $R_{1/2}\simeq6.69$ fm.}
 \label{fig:omegaccc-waves}
\end{figure}

At larger $\alpha$, the scalar attraction draws these states into the nuclear region, as illustrated by Fig.~\ref{fig:omega-waves}(b). At $\alpha=3$ in ${}^{12}\mathrm{C}$, the rms radii of all four states lie between 1.53 and 2.93 fm. A smooth central minimum can then give an approximately harmonic-oscillator pattern. The calculated $1p-1s$ and $1d-1p$ spacings are 17.07 and 15.85 MeV, while the $2s$ and $1d$ levels differ by 1.64 MeV. The grouping has thus changed from approximate $2s$--$1p$ pairing to approximate $2s$--$1d$ pairing. The finite range and diffuse surface of the potential prevent an exact oscillator degeneracy.

Increasing the nuclear charge provides another route to interior localization. In ${}^{208}\mathrm{Pb}$, even at $\alpha=1$, the attractive Coulomb field places the four $\Omega$ states predominantly inside $R_{1/2}=6.69$ fm [Fig.~\ref{fig:omega-waves}(c)]; the probability in this
region exceeds 93\% for each state. The $1p-1s$ and
$1d-1p$ spacings are 3.47 and 3.66 MeV, and the $2s$--$1d$ splitting is only 0.28 MeV. This oscillator-like grouping is consistent with the approximately quadratic Coulomb potential inside a finite charge distribution, together with the scalar attraction. Along the nuclei studied here, the increasing charge and the larger spatial extent both influence the spectrum. The trend is therefore not attributed to the mass number alone. The nuclear core can also rearrange in light cluster systems: using HAL QCD two body input, Zhou and Liu~\cite{ZhouLiu2026} predicted a compact ${}^{9}_{\Omega}\mathrm{Be}$ system, with their $0_1^+$ level at approximately $-46.4$ MeV relative to the $\alpha+\alpha+\Omega$ threshold and a contracted $\alpha\alpha$ core. Their interaction input and dynamical cluster treatment differ from the present fixed core description, motivating a systematic comparison in future work.

For $\Occc$, the scalar attraction must instead overcome Coulomb repulsion. Binding therefore requires sufficient scalar attraction, and its strength alone does not determine which nuclear core gives the deepest state. In lighter nuclei, the limited spatial extent of the attractive potential and the associated localization kinetic energy restrict the binding. In heavier nuclei, the increasing Coulomb repulsion offsets the gain from the larger attractive region. This competition favors a core between the lightest and heaviest nuclei considered. Among the six cores studied here, ${}^{16}\mathrm{O}$ gives the deepest $1s$ state at each sampled cutoff that supports binding, namely $\alpha=1.5$, 2, 2.5, and 3. At $\alpha=3$, for example, $E_{1s}=-44.02$ MeV in ${}^{16}\mathrm{O}$, compared with $-41.46$ MeV in ${}^{12}\mathrm{C}$ and $-17.72$ MeV in ${}^{208}\mathrm{Pb}$. These results motivate a broader search for the nuclear cores most favorable to $\Occc$ binding.

When all four selected orbitals are bound in all six cores, as at $\alpha=2.5$ and 3, the light  and heavy nucleus spectra exhibit different patterns. Taking $\alpha=3$ as an example, the strong attraction in ${}^{12}\mathrm{C}$ supports compact wave functions [Fig.~\ref{fig:omegaccc-waves}(a)] with rms radii of 1.25--1.96 fm. The $1p-1s$ and $1d-1p$ spacings are 11.10 and 11.31 MeV, respectively, and the $2s$--$1d$ splitting is 0.29 MeV. These low-lying levels again follow the approximate oscillator grouping.

The heavy-nucleus spectrum has a different organization. For $\Occc$--${}^{208}\mathrm{Pb}$ at $\alpha=3$, the total potential has its minimum at $r_{\min}\simeq4.92$ fm, away from the center [Fig.~\ref{fig:potentials}(d)]. The $1s$, $1p$, and $1d$ wave functions in Fig.~\ref{fig:omegaccc-waves}(b) occupy nearly the same radial region, with mean radii of 4.16, 4.34, and 4.51 fm. At such radii the centrifugal energy is relatively small, allowing these levels to form a compressed group. Indeed, $E_{1p}-E_{1s}=0.58$ MeV and $E_{1d}-E_{1s}=1.56$ MeV, whereas the radial excitation costs $E_{2s}-E_{1s}=3.91$ MeV. The $2s$ wave function has one radial node and samples a broader radial region extending farther toward the nuclear center. Relative to the $1s$ state, this redistribution increases both the radial kinetic energy and the mean potential energy. Its larger excitation energy therefore reflects the cost of changing the radial structure, whereas the $1p$ and $1d$ splittings mainly arise from the comparatively small centrifugal contribution.

For both baryons, changing $\alpha$ alters the scalar attraction relative to a fixed Coulomb field and thereby changes the regions sampled by the wave functions. Consequently, the cutoff dependence cannot generally be represented as a common shift of all levels. For example, the $\Omega$--${}^{12}\mathrm{C}$ $1p-1s$ spacing increases from 2.35 to 17.07 MeV between $\alpha=1$ and 3. The approximate cutoff stability of excitation energies found for neutral $\chi_{cJ}$ systems~\cite{Gao2026} therefore does not extend uniformly to these charged baryons. The observed level patterns depend on both the strength of the scalar attraction and the charge and geometry of the nuclear core. The cosh potential representation of this interplay is left to the following section.

\section{Extension of the cosh potential to charged baryons}
\label{sec:cosh}

The QMC based study of neutral $\chi_{cJ}(1P)$--nucleus systems~\cite{Gao2026} showed that a cosh potential with common radius and surface diffuseness parameters describes the calculated energies and excitation gaps at the few-MeV and sub-MeV levels, respectively. The dependence on the charmonium state and cutoff was carried mainly by the potential depth. This motivates extending the same geometry to $\Omega$ and $\Occc$, whose spectra also depend on the Coulomb interaction. We first test the scalar-only $\Omega$ spectrum by switching off Coulomb. For either baryon, we write
\begin{equation}
 \begin{aligned}
 S_{H,\mathrm{cosh}}^A(r;\alpha)
 &=-V_0^H(\alpha)f_A(r),\\
 f_A(r)&=\frac{1+\cosh(R_A/a_s)}
                   {\cosh(r/a_s)+\cosh(R_A/a_s)},
 \end{aligned}
 \label{eq:cosh-scalar}
\end{equation}
with $R_A=r_0A^{1/3}$, $r_0=1.05$ fm, and $a_s=0.53$ fm. The depth is fixed by $V_0^H(\alpha)=-\Delta M_H(\rho_0;\alpha)$, using the nuclear matter results in Table~\ref{tab:mass-shifts}. The geometry is unchanged from Ref.~\cite{Gao2026}; no parameter is fitted to the present finite nucleus spectra. Equation~\eqref{eq:schrodinger} is solved with the appropriate baryon and nuclear reduced masses.

We compare the cosh and original potential spectra using the root-mean-square (RMS) energy deviation,
\begin{equation}
 \mathrm{RMS}(E)=
 \left[\frac{1}{N}\sum_{i=1}^{N}
       \bigl(E_i^{\mathrm{cosh}}-E_i^{\mathrm{ref}}\bigr)^2
 \right]^{1/2},
 \label{eq:cosh-rms}
\end{equation}
where $N$ counts the levels bound in both calculations across the cores, orbitals, and cutoffs considered for each baryon. The superscripts distinguish the cosh and reference results. For the excitation energies $\Delta E_{n\ell}=E_{n\ell}-E_{1s}$ within each core and cutoff, $\mathrm{RMS}(\Delta E)$ is evaluated in the same way, with $N$ counting the common excitation energies. Both the excited state and its $1s$ reference must be bound in both calculations. Missing and additional bound states are counted separately. Energies, energy differences, and RMS deviations are reported to two decimal places. Bound state classifications follow the original numerical results.

For the scalar-only $\Omega$ calculation, this prescription retains all 98 reference bound levels among the six cores, four orbitals, and five cutoffs considered. The RMS deviations of the energies and excitation gaps are 2.26 and 0.63 MeV, respectively. One additional shallow $1p$ state appears in ${}^{16}\mathrm{O}$ at $\alpha=1.5$, with $E=-0.11$ MeV. Thus the original geometry provides a useful approximation to the scalar spectrum, although its threshold classification is not exact. The individual energies are listed in Appendix~\ref{app:spectra}.

To include charge without introducing a new radius parameter, we assign a normalized charge distribution the same shape $f_A(r)$ and match its mean-square radius to that of a uniform sphere. The cosh profile gives
\begin{equation}
 \begin{aligned}
 \langle r^2\rangle_f
 &=\frac{\int_0^\infty r^4 f_A(r)\,\dd r}
          {\int_0^\infty r^2 f_A(r)\,\dd r}
   =\frac35 R_A^2+\frac75\pi^2a_s^2,\\
 R_C&=\sqrt{\frac53\langle r^2\rangle_f}
     =\sqrt{R_A^2+\frac73\pi^2a_s^2}.
 \end{aligned}
 \label{eq:cosh-charge-radius}
\end{equation}
Here $R_C$ is the sphere boundary. This matches the second moment of the assumed charge distribution; it does not use the bound state wave functions or fit the QMC charge radii. Sharing the scalar and charge geometry is a modeling assumption. The resulting Coulomb contribution is
\begin{equation}
 V_{C,H}^{A,\mathrm{cosh}}(r)=Q_HZ\alpha_{\rm em}
 \begin{cases}
  \displaystyle\frac{3-r^2/R_C^2}{2R_C}, & r\le R_C,\\[3pt]
  \displaystyle\frac1r, & r>R_C,
 \end{cases}
 \label{eq:cosh-coulomb}
\end{equation}
in natural units, where $Z$ is the proton number of the nuclear core
and $\alpha_{\rm em}\simeq 1/137$ is the fine structure constant.
The baryon charges in units of $e$ are $Q_\Omega=-1$ and $Q_{\Occc}=+2$. We solve Eq.~\eqref{eq:schrodinger} with $U_H^A=S_{H,\mathrm{cosh}}^A+V_{C,H}^{A,\mathrm{cosh}}$, including the change in each wave function caused by the Coulomb field.

Table~\ref{tab:cosh-accuracy} summarizes these comparisons. For charged $\Omega$, all 120 reference levels remain bound, with an energy RMS of 2.04 MeV and a gap RMS of 0.61 MeV. The largest energy deviation is 6.15 MeV for ${}^{40}\mathrm{Ca}$ $2s$ at $\alpha=3$.

\begin{table}[t]
 \caption{Accuracy of the cosh representation. RMS deviations
are in MeV. The second column lists the numbers of common and reference bound levels, respectively. The last column counts missed/additional bound states among the 120 combinations examined in each row. Gap RMS values use 68, 90, and 45 common excitation gaps, respectively. Only the first row omits Coulomb.}
 \label{tab:cosh-accuracy}
 \centering
 \setlength{\tabcolsep}{2pt}
 \begin{tabular*}{\columnwidth}{@{\extracolsep{\fill}}ccccc}
  \toprule\toprule
  System & Common/Ref. & $\mathrm{RMS}(E)$ & $\mathrm{RMS}(\Delta E)$ & Miss/add \\
  \midrule
  $\Omega$, $V_C=0$ & $98/98$ & 2.26 & 0.63 & $0/1$ \\
  $\Omega$ & $120/120$ & 2.04 & 0.61 & $0/0$ \\
  $\Occc$ & $63/66$ & 2.37 & 0.48 & $3/0$ \\
  \bottomrule\bottomrule
 \end{tabular*}
\end{table}

For $\Occc$, replacing only the baryon mass, scalar depths, and charge preserves 63 of the 66 reference bound levels. Their energy and gap RMS deviations are 2.37 and 0.48 MeV. The three missed states are $2s$ in ${}^{90}\mathrm{Zr}$ at $\alpha=2$ and in ${}^{197}\mathrm{Au}$ and ${}^{208}\mathrm{Pb}$ at $\alpha=2.5$, whose reference energies are $-1.11$, $-1.84$, and $-1.86$ MeV. All 18 reference $1s$ levels are retained. Applying Eq.~\eqref{eq:cosh-rms} to these levels defines the ground-state RMS residual $\epsilon_g\simeq2.16$ MeV, where $g$ denotes the ground state. Tables~\ref{tab:app-omega-charged} and~\ref{tab:app-ccc-charged} place the original charged spectra and the cosh results side by side.

The remaining discrepancies reflect the radial information lost when a single nuclear matter depth and fixed geometry replace the nucleus dependent scalar profiles. For example, the original $\Occc$ scalar potential in ${}^{40}\mathrm{Ca}$ at $\alpha=3$ reaches about $-74.70$ MeV at the center, whereas Eq.~\eqref{eq:cosh-scalar} gives $-65.02$ MeV. Keeping the original scalar potentials and replacing only their Coulomb fields by Eq.~\eqref{eq:cosh-coulomb} retains all 66 $\Occc$ bound levels, with an energy RMS of 0.83 MeV. This control identifies the scalar profile approximation as the main limitation, with Coulomb repulsion making shallow levels more sensitive to it.

The smaller gap residuals indicate partial cancellation of energy shifts within a given nucleus and cutoff. They do not imply that the gaps are independent of $\alpha$. The cosh form with the matched Coulomb field therefore provides a compact description of the overall binding and level spacings, while individual near-threshold states and small energy differences between different nuclei require the original potentials.

The compact potential also permits a broader exploration of the cores favorable to the deepest $\Occc$ ground state. We retain the geometry and matched Coulomb field introduced above, but allow the common scalar depth to vary over $0\leq V_0\leq70$ MeV. At each depth, the minimum is taken over the negative energy $1s$ solutions in the specified core set; no optimum is assigned when none is bound. This continuous scan tracks the preferred core as a function of the scalar attraction strength. The sampled cutoff values enter through their corresponding depths $V_0(\alpha)$.

\begin{figure}[tp]
 \centering
 \includegraphics[width=\columnwidth]{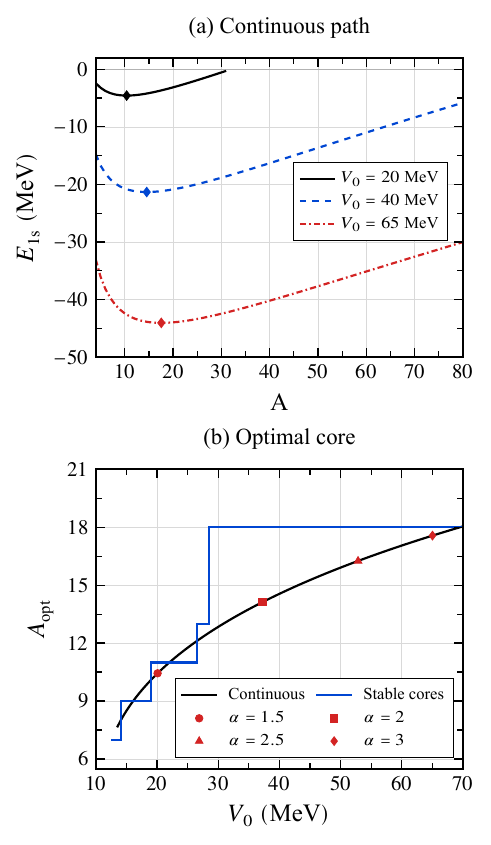}
 \caption{Exploration of the deepest $\Occc$ ground
state with the cosh scalar potential and RMS-matched Coulomb field. (a) Ground-state energies along the continuous $N=Z$ path at $V_0=20$, 40, and 65 MeV; diamonds mark the minima. Only bound portions of the curves are shown. (b) Minimizing mass numbers for the continuous path (black) and the nominal best stable cores (blue). From left to right, the five plateaus correspond to ${}^{7}\mathrm{Li}$, ${}^{9}\mathrm{Be}$, ${}^{11}\mathrm{B}$, ${}^{13}\mathrm{C}$, and ${}^{18}\mathrm{O}$. Their transitions occur at $V_0\simeq14.14$, 19.01, 26.54, and 28.52 MeV; adjacent cores share the minimum at a transition. The first stable core binds above approximately 12.29 MeV. Red symbols mark the continuous optima at the nuclear-matter depths for the indicated cutoffs. The Li, Be, and B plateaus are extrapolations below $A=12$, the lightest core in the six-nucleus comparison.}
 \label{fig:cosh-optimum}
\end{figure}

We first use the ideal continuous path $4\leq A\leq300$, with $Z=A/2$ and $M_A=A\,u$, where $u=931.49$ MeV. Its noninteger minimizing mass numbers $A_{\mathrm{opt}}$ do not identify physical nuclides. In a small core, the limited range of the scalar attraction entails a large kinetic-energy cost of localization. Increasing $A$ enlarges the attractive region and increases the reduced mass, but also strengthens Coulomb repulsion along this charge path. Their competition produces the finite minima in Fig.~\ref{fig:cosh-optimum}(a). The first bound minimum appears near $V_0=13.33$ MeV. With increasing depth, the optimum moves smoothly to larger cores, from $A_{\mathrm{opt}}=10.40$ at 20 MeV to 18.05 at 70 MeV [Fig.~\ref{fig:cosh-optimum}(b)].

We separately examine the 250 ground-state nuclides marked as stable in NUBASE2020~\cite{Kondev2021}, with $A\geq4$, using their actual integer $A$ and $Z$ and nuclear masses obtained from the evaluated atomic masses. This discrete scan yields the staircase in Fig.~\ref{fig:cosh-optimum}(b). Within the scanned interval, ${}^{18}\mathrm{O}$ is the nominal optimum above $V_0\simeq28.52$ MeV. At fixed $Z$, adding neutrons increases the radius and reduced mass without increasing the total Coulomb charge, which can favor ${}^{18}\mathrm{O}$ over ${}^{16}\mathrm{O}$ in this approximation. The earlier heavy-quarkonium benchmarks in Ref.~\cite{Gao2026} gave a reasonable description of the ${}^{48}\mathrm{Ca}$ spectra without refitting the common geometry, supporting its application beyond the calibration nuclei. We therefore extend the same scalar depth and radius prescription to additional isotopes, whose individual $\Occc$ QMC scalar potentials have not yet been tested. Predictions for optimal cores with $A<12$ additionally extrapolate below the lightest core in the six-nucleus comparison, ${}^{12}\mathrm{C}$.

Table~\ref{tab:cosh-optimal-cores} gives the best and next best stable cores at the sampled cutoffs. At $\alpha=1.5$, the unrestricted scan favors ${}^{11}\mathrm{B}$; restricting the candidates to $A\geq12$ instead gives ${}^{13}\mathrm{C}$. For $\alpha=2$, 2.5, and 3, the nominal winner is ${}^{18}\mathrm{O}$, but its separation from the next best core is only 0.28--0.45 MeV. Deeper binding therefore does not by itself make the isotope ranking more secure.

\begin{table}[t]
 \caption{Nominal best and next best stable cores in the
cosh model at the depths corresponding to the sampled cutoffs. $E_{\min}$ and $\delta E=E_{\mathrm{next}}-E_{\min}$ are in MeV. A dash denotes the absence of a bound optimum in the stable-core set. The $\alpha=1.5$ optimum is lighter than the lightest reference core; imposing $A\geq12$ changes it to ${}^{13}\mathrm{C}$.}
 \label{tab:cosh-optimal-cores}
 \centering
 \setlength{\tabcolsep}{3pt}
 \begin{tabular*}{\columnwidth}{@{\extracolsep{\fill}}ccccc}
  \toprule\toprule
  $\alpha$ & Best core & $E_{\min}$ & Next core & $\delta E$ \\
  \midrule
  1 & \textemdash & \textemdash & \textemdash & \textemdash \\
  1.5 & ${}^{11}\mathrm{B}$ & $-5.21$ & ${}^{9}\mathrm{Be}$ & 0.04 \\
  2 & ${}^{18}\mathrm{O}$ & $-19.74$ & ${}^{15}\mathrm{N}$ & 0.28 \\
  2.5 & ${}^{18}\mathrm{O}$ & $-33.93$ & ${}^{22}\mathrm{Ne}$ & 0.45 \\
  3 & ${}^{18}\mathrm{O}$ & $-45.17$ & ${}^{22}\mathrm{Ne}$ & 0.31 \\
  \bottomrule\bottomrule
 \end{tabular*}
\end{table}

\begin{figure}[tp]
 \centering
 \includegraphics[width=\columnwidth]{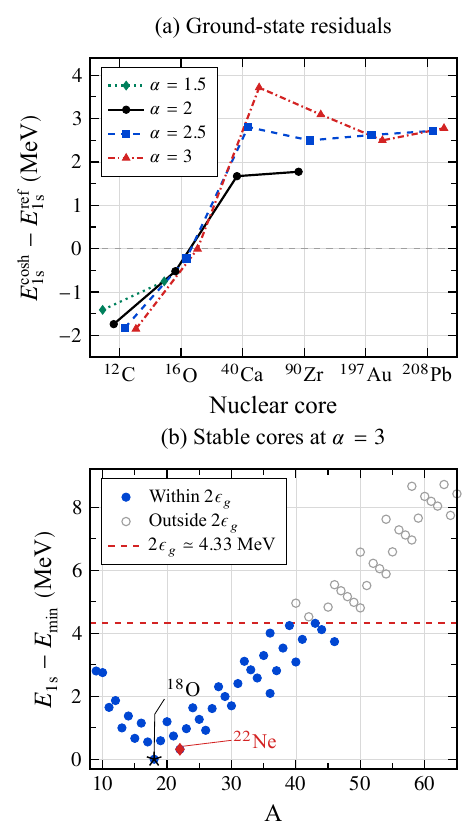}
 \caption{Ground-state reproduction and ranking sensitivity
for $\Occc$. (a) Residuals $E_{1s}^{\mathrm{cosh}}-E_{1s}^{\mathrm{ref}}$ relative to the original potentials at the four cutoffs supporting reference binding. Points are included only when the $1s$ level is bound in both calculations; the $\alpha=1.5$ series therefore contains only ${}^{12}\mathrm{C}$ and ${}^{16}\mathrm{O}$. (b) Energy deficits of the displayed stable cores from the nominal minimum at $\alpha=3$. Blue filled and gray open circles lie within and outside $2\epsilon_g$, respectively. The black star identifies ${}^{18}\mathrm{O}$, and the red diamond identifies the next best core, ${}^{22}\mathrm{Ne}$. The dashed line marks the assumed sensitivity threshold in Eq.~\eqref{eq:cosh-ranking-sensitivity}, rather than a statistical confidence limit.}
 \label{fig:cosh-ranking-sensitivity}
\end{figure}

Resolving these small differences requires greater accuracy than reproducing the overall binding scale. The ground state RMS residual $\epsilon_g$ defined above pools the 18 common bound $1s$ levels across $\alpha=1.5$, 2, 2.5, and 3, with equal weight assigned to each level. Figure~\ref{fig:cosh-ranking-sensitivity}(a) shows that the deviations depend on the core: carbon is overbound, whereas calcium is underbound. At $\alpha=3$, the original ${}^{16}\mathrm{O}$--${}^{40}\mathrm{Ca}$ ground-state separation is only 0.09 MeV, compared with 3.81 MeV in the cosh representation. A common energy offset would preserve the ranking, but these differences need not cancel.

To illustrate the resulting ranking sensitivity, suppose that each ground-state energy receives a shift $\delta E_i$ satisfying $|\delta E_i|\leq\epsilon_g$, without requiring a common shift across cores. A candidate can become at least as deeply bound as the nominal winner if
\begin{equation}
 E_{1s,i}^{\mathrm{cosh}}-E_{\min}^{\mathrm{cosh}}
 \leq2\epsilon_g.
 \label{eq:cosh-ranking-sensitivity}
\end{equation}
The factor two allows opposite shifts of the two energies. Here the benchmark RMS supplies a representative sensitivity scale, not a proven bound on each residual or a statistical confidence interval. Figure~\ref{fig:cosh-ranking-sensitivity}(b) uses $\alpha=3$ as an illustrative strong-attraction case. At $2\epsilon_g\simeq4.33$ MeV, 37 stable candidate nuclei satisfy Eq.~\eqref{eq:cosh-ranking-sensitivity}, with mass numbers ranging from $A=9$ to 46. The nominal ${}^{18}\mathrm{O}$--${}^{22}\mathrm{Ne}$ difference of 0.31 MeV is much smaller than this scale. These variations assess only the cosh representation relative to the original potentials; its benchmark accuracy does not reliably resolve the small energy differences between the leading candidates. More importantly, the QMC based reference treatment employs the local density approximation and fixed spherical nuclear cores, neglecting nuclear deformation and the possible core response to the added $\Occc$. These limitations may affect light nucleus predictions and prevent a reliable identification of a unique physical optimum within the present framework.

\section{Summary and outlook}
\label{sec:summary}

We have investigated the nuclear bound-state spectra of $\Omega$ and $\Occc$
in a QMC-based hadronic-loop approach. The in-medium modifications of the
intermediate hadrons generate scalar attraction, which is mapped to finite
nuclei through the local density approximation and combined with the Coulomb
field to obtain the bound-state energies and radial wave functions. The
results depend on the free cutoff parameter $\alpha$ and, for $\Occc$, on the
assumed $SU(4)$ relations among the hadronic couplings.

For $\Omega$, the scalar and Coulomb attractions act together to promote
nuclear localization. In light nuclear cores at weak scalar coupling, the
shallow excited states extend well beyond the nucleus and exhibit an
approximately hydrogen-like $2s$--$1p$ grouping. Stronger scalar attraction
or a larger nuclear charge draws the low-lying states inward, leading to an
approximately oscillator-like $2s$--$1d$ grouping in the cases examined. These
patterns reflect the different radial regions sampled by the states.

For $\Occc$, the scalar attraction competes with the Coulomb repulsion, so
binding requires sufficient scalar attraction and does not improve
monotonically with nuclear size. When the selected orbitals are all bound, the
light-core examples exhibit approximately oscillator-like spectra. In the
heavy-core examples, an off-center potential minimum favors similar radial
distributions for the nodeless $1s$, $1p$, and $1d$ states, producing a
compressed group of low-lying levels. The $2s$ radial excitation entails
larger kinetic- and potential-energy costs and remains more widely separated.
For both baryons, the excitation gaps depend on the balance between scalar
and Coulomb effects; their approximate stability with respect to the cutoff is
not a general property.

The cosh representation with the previously adopted geometry first reproduces
the scalar-only $\Omega$ spectrum and, with an RMS-matched Coulomb field,
extends to both charged systems. The energy RMS deviations are of the order of
a few MeV, while those for excitation gaps remain below 1 MeV; individual
near-threshold states are more sensitive to the approximation. A continuous
scan of the scalar depth reveals preferred cores determined by the competition
between the range of the attraction, the localization kinetic energy, and the
Coulomb repulsion. The scan suggests that relatively light nuclei favor the
deepest $\Occc$ binding within the present approximations. In the illustrative
cosh sensitivity analysis at $\alpha=3$, 37 stable candidate cores spanning
$9\leq A\leq 46$ lie within 4.33 MeV of the nominal minimum. A unique optimal
nucleus cannot be reliably identified, given the limited cosh accuracy and the
use of the local density approximation and fixed spherical nuclear cores in
the QMC-based reference calculation.

Future $\Omega N$ and $\Omega$-atom studies at J-PARC~\cite{Aoki2021P85},
together with the multicharm program of ALICE 3~\cite{ALICE2022LoI}, may
provide experimental input for exploring these systems. Assessing their
observability requires formation rates and conversion and decay widths beyond
the present real-potential calculation. Applications to heavy-ion collisions
must also account for finite-temperature in-medium effects.

\begin{acknowledgments}
This work is supported by the Natural Science Foundation of Gansu Province (No. 26RCKA012 and No. 25JRRA799), the National Natural Science Foundation of China under Grants No. 12335001 and No. 12247101, the ``111 Center'' under Grant No. B20063, the Fundamental Research Funds for the Central Universities (lzujbky-2023-stlt01), and Lanzhou City High-Level Talent Funding. T.-L. G. is supported by the Gansu Province Postgraduate Innovation Star Program No. 2026CXZX-042. The authors used ChatGPT for language editing and for assistance with the numerical code, reviewed the output, and take responsibility for the content.
\end{acknowledgments}

\appendix
\begin{table*}[!t]
 \caption{Scalar-only $\Omega$ energies with the Coulomb interaction switched off in both calculations. The reference uses the original scalar potentials; the cosh result uses Eq.~\eqref{eq:cosh-scalar}. All energies are in MeV and are rounded to two decimal places. Each cutoff has paired reference (Ref.) and cosh columns. A dash denotes no bound state in the specified orbital; $\dagger$ marks a difference in bound-state classification.}
 \label{tab:app-omega-scalar}
 \centering
 \setlength{\tabcolsep}{3pt}
 \begin{tabular*}{\textwidth}{@{\extracolsep{\fill}}cc*{10}{c}}
  \toprule\toprule
  Core & State & \multicolumn{2}{c}{$\alpha=1$} & \multicolumn{2}{c}{$\alpha=1.5$} & \multicolumn{2}{c}{$\alpha=2$} & \multicolumn{2}{c}{$\alpha=2.5$} & \multicolumn{2}{c}{$\alpha=3$} \\
  \cmidrule(lr){3-4} \cmidrule(lr){5-6} \cmidrule(lr){7-8} \cmidrule(lr){9-10} \cmidrule(lr){11-12}
  & & Ref. & Cosh & Ref. & Cosh & Ref. & Cosh & Ref. & Cosh & Ref. & Cosh \\
  \midrule
  ${}^{12}\mathrm{C}$ & $1s$ & $-0.13$ & $-0.61$ & $-4.79$ & $-5.84$ & $-14.56$ & $-15.52$ & $-25.85$ & $-26.58$ & $-36.24$ & $-36.75$ \\
   & $1p$ & \textemdash & \textemdash & \textemdash & \textemdash & $-3.12$ & $-4.08$ & $-11.23$ & $-12.19$ & $-19.52$ & $-20.37$ \\
   & $1d$ & \textemdash & \textemdash & \textemdash & \textemdash & \textemdash & \textemdash & \textemdash & \textemdash & $-4.12$ & $-4.58$ \\
   & $2s$ & \textemdash & \textemdash & \textemdash & \textemdash & \textemdash & \textemdash & $-1.41$ & $-1.23$ & $-5.87$ & $-5.53$ \\
  \midrule
  ${}^{16}\mathrm{O}$ & $1s$ & $-0.69$ & $-1.11$ & $-7.00$ & $-7.19$ & $-18.16$ & $-17.63$ & $-30.53$ & $-29.28$ & $-41.72$ & $-39.89$ \\
   & $1p$ & \textemdash & \textemdash & \textemdash & $-0.11$$^{\dagger}$ & $-7.13$ & $-6.85$ & $-17.27$ & $-16.13$ & $-27.01$ & $-25.13$ \\
   & $1d$ & \textemdash & \textemdash & \textemdash & \textemdash & \textemdash & \textemdash & $-4.20$ & $-3.25$ & $-11.84$ & $-10.10$ \\
   & $2s$ & \textemdash & \textemdash & \textemdash & \textemdash & $-0.08$ & $-0.01$ & $-4.65$ & $-3.80$ & $-11.28$ & $-9.80$ \\
  \midrule
  ${}^{40}\mathrm{Ca}$ & $1s$ & $-3.38$ & $-3.06$ & $-12.54$ & $-11.16$ & $-26.06$ & $-23.35$ & $-40.34$ & $-36.34$ & $-53.04$ & $-47.93$ \\
   & $1p$ & \textemdash & \textemdash & $-5.98$ & $-5.07$ & $-17.88$ & $-15.57$ & $-31.02$ & $-27.43$ & $-42.90$ & $-38.24$ \\
   & $1d$ & \textemdash & \textemdash & \textemdash & \textemdash & $-9.32$ & $-7.14$ & $-21.12$ & $-17.52$ & $-32.10$ & $-27.37$ \\
   & $2s$ & \textemdash & \textemdash & $-0.44$ & $-0.02$ & $-8.72$ & $-6.07$ & $-20.15$ & $-15.73$ & $-31.01$ & $-25.19$ \\
  \midrule
  ${}^{90}\mathrm{Zr}$ & $1s$ & $-5.35$ & $-4.71$ & $-15.36$ & $-13.85$ & $-29.44$ & $-26.90$ & $-44.07$ & $-40.55$ & $-56.98$ & $-52.60$ \\
   & $1p$ & $-1.97$ & $-1.58$ & $-10.93$ & $-9.51$ & $-24.28$ & $-21.73$ & $-38.39$ & $-34.81$ & $-50.93$ & $-46.48$ \\
   & $1d$ & \textemdash & \textemdash & $-6.02$ & $-4.59$ & $-18.45$ & $-15.71$ & $-31.94$ & $-28.09$ & $-44.04$ & $-39.30$ \\
   & $2s$ & \textemdash & \textemdash & $-5.20$ & $-3.56$ & $-17.18$ & $-14.03$ & $-30.46$ & $-26.09$ & $-42.44$ & $-37.11$ \\
  \midrule
  ${}^{197}\mathrm{Au}$ & $1s$ & $-6.07$ & $-5.97$ & $-15.92$ & $-15.65$ & $-29.55$ & $-29.13$ & $-43.62$ & $-43.09$ & $-55.99$ & $-55.37$ \\
   & $1p$ & $-3.98$ & $-3.67$ & $-13.48$ & $-12.78$ & $-26.91$ & $-25.86$ & $-40.87$ & $-39.55$ & $-53.16$ & $-51.65$ \\
   & $1d$ & $-1.53$ & $-1.07$ & $-10.46$ & $-9.35$ & $-23.58$ & $-21.91$ & $-37.33$ & $-35.27$ & $-49.49$ & $-47.15$ \\
   & $2s$ & $-0.87$ & $-0.57$ & $-9.14$ & $-8.17$ & $-21.92$ & $-20.47$ & $-35.43$ & $-33.68$ & $-47.41$ & $-45.46$ \\
  \midrule
  ${}^{208}\mathrm{Pb}$ & $1s$ & $-6.40$ & $-6.04$ & $-16.53$ & $-15.75$ & $-30.51$ & $-29.25$ & $-44.95$ & $-43.22$ & $-57.65$ & $-55.51$ \\
   & $1p$ & $-4.19$ & $-3.80$ & $-13.87$ & $-12.96$ & $-27.53$ & $-26.08$ & $-41.73$ & $-39.81$ & $-54.24$ & $-51.92$ \\
   & $1d$ & $-1.70$ & $-1.25$ & $-10.76$ & $-9.63$ & $-24.04$ & $-22.26$ & $-37.97$ & $-35.67$ & $-50.30$ & $-47.57$ \\
   & $2s$ & $-1.22$ & $-0.71$ & $-9.87$ & $-8.47$ & $-23.02$ & $-20.85$ & $-36.91$ & $-34.11$ & $-49.23$ & $-45.92$ \\
  \bottomrule\bottomrule
 \end{tabular*}
\end{table*}

\section{Reference spectra and cosh comparisons}
\label{app:spectra}

The complete numerical comparisons for the selected $1s$, $1p$, $1d$, and $2s$ orbitals are given below. Table~\ref{tab:app-omega-scalar} tests the scalar-only $\Omega$ spectrum, while Tables~\ref{tab:app-omega-charged} and~\ref{tab:app-ccc-charged} give the charged spectra plotted in Fig.~\ref{fig:omega-omegaccc-level-spectra} together with their cosh approximations. All cosh entries use $r_0=1.05$ fm and $a_s=0.53$ fm, with the depths specified in Sec.~\ref{sec:cosh}. The charged calculations employ only the mean-square-radius prescription of Eq.~\eqref{eq:cosh-charge-radius}. Negative energies indicate binding. Nonnegative finite-box eigenvalues are represented by dashes and are not assigned physical resonance energies.

\begin{table*}[!t]
 \caption{Charged $\Omega$ energies. The reference entries are the numerical values underlying Fig.~\ref{fig:omega-omegaccc-level-spectra}(a), including the original Coulomb field. The cosh entries use Eqs.~\eqref{eq:cosh-scalar} and~\eqref{eq:cosh-coulomb}. All energies are in MeV and are rounded to two decimal places. Each cutoff has paired reference (Ref.) and cosh columns. A dash denotes no bound state in the specified orbital; $\dagger$ marks a difference in bound-state classification.}
 \label{tab:app-omega-charged}
 \centering
 \setlength{\tabcolsep}{3pt}
 \begin{tabular*}{\textwidth}{@{\extracolsep{\fill}}cc*{10}{c}}
  \toprule\toprule
  Core & State & \multicolumn{2}{c}{$\alpha=1$} & \multicolumn{2}{c}{$\alpha=1.5$} & \multicolumn{2}{c}{$\alpha=2$} & \multicolumn{2}{c}{$\alpha=2.5$} & \multicolumn{2}{c}{$\alpha=3$} \\
  \cmidrule(lr){3-4} \cmidrule(lr){5-6} \cmidrule(lr){7-8} \cmidrule(lr){9-10} \cmidrule(lr){11-12}
  & & Ref. & Cosh & Ref. & Cosh & Ref. & Cosh & Ref. & Cosh & Ref. & Cosh \\
  \midrule
  ${}^{12}\mathrm{C}$ & $1s$ & $-2.72$ & $-3.31$ & $-8.28$ & $-9.04$ & $-18.31$ & $-18.87$ & $-29.71$ & $-30.00$ & $-40.17$ & $-40.20$ \\
   & $1p$ & $-0.37$ & $-0.39$ & $-0.91$ & $-1.39$ & $-6.27$ & $-7.10$ & $-14.68$ & $-15.40$ & $-23.10$ & $-23.66$ \\
   & $1d$ & $-0.15$ & $-0.16$ & $-0.15$ & $-0.16$ & $-0.15$ & $-0.16$ & $-1.71$ & $-2.05$ & $-7.25$ & $-7.60$ \\
   & $2s$ & $-0.43$ & $-0.45$ & $-0.57$ & $-0.57$ & $-1.06$ & $-1.02$ & $-3.99$ & $-3.72$ & $-8.89$ & $-8.42$ \\
  \midrule
  ${}^{16}\mathrm{O}$ & $1s$ & $-4.55$ & $-4.77$ & $-11.56$ & $-11.33$ & $-22.96$ & $-21.93$ & $-35.43$ & $-33.65$ & $-46.68$ & $-44.29$ \\
   & $1p$ & $-0.79$ & $-0.89$ & $-3.26$ & $-3.46$ & $-11.41$ & $-10.82$ & $-21.79$ & $-20.27$ & $-31.63$ & $-29.34$ \\
   & $1d$ & $-0.28$ & $-0.29$ & $-0.29$ & $-0.29$ & $-1.20$ & $-0.97$ & $-8.21$ & $-7.04$ & $-16.09$ & $-14.07$ \\
   & $2s$ & $-0.74$ & $-0.74$ & $-1.00$ & $-0.98$ & $-2.80$ & $-2.47$ & $-8.48$ & $-7.40$ & $-15.44$ & $-13.68$ \\
  \midrule
  ${}^{40}\mathrm{Ca}$ & $1s$ & $-12.28$ & $-11.48$ & $-21.89$ & $-20.02$ & $-35.61$ & $-32.39$ & $-50.01$ & $-45.48$ & $-62.78$ & $-57.12$ \\
   & $1p$ & $-6.75$ & $-6.61$ & $-14.53$ & $-13.32$ & $-26.84$ & $-24.21$ & $-40.17$ & $-36.23$ & $-52.15$ & $-47.12$ \\
   & $1d$ & $-2.49$ & $-2.55$ & $-7.28$ & $-6.48$ & $-17.70$ & $-15.33$ & $-29.81$ & $-25.99$ & $-40.94$ & $-35.97$ \\
   & $2s$ & $-3.55$ & $-3.38$ & $-7.48$ & $-6.36$ & $-17.14$ & $-14.16$ & $-28.95$ & $-24.22$ & $-39.99$ & $-33.84$ \\
  \midrule
  ${}^{90}\mathrm{Zr}$ & $1s$ & $-20.04$ & $-19.06$ & $-30.39$ & $-28.58$ & $-44.66$ & $-41.83$ & $-59.41$ & $-55.59$ & $-72.40$ & $-67.70$ \\
   & $1p$ & $-15.45$ & $-14.85$ & $-25.10$ & $-23.54$ & $-38.79$ & $-36.11$ & $-53.08$ & $-49.36$ & $-65.73$ & $-61.13$ \\
   & $1d$ & $-10.76$ & $-10.42$ & $-19.40$ & $-17.95$ & $-32.33$ & $-29.60$ & $-46.07$ & $-42.23$ & $-58.31$ & $-53.56$ \\
   & $2s$ & $-10.63$ & $-10.09$ & $-18.70$ & $-16.94$ & $-31.29$ & $-28.12$ & $-44.83$ & $-40.45$ & $-56.95$ & $-51.61$ \\
  \midrule
  ${}^{197}\mathrm{Au}$ & $1s$ & $-28.68$ & $-29.36$ & $-38.74$ & $-39.37$ & $-52.49$ & $-53.05$ & $-66.63$ & $-67.12$ & $-79.03$ & $-79.47$ \\
   & $1p$ & $-25.44$ & $-25.86$ & $-35.38$ & $-35.56$ & $-49.05$ & $-48.96$ & $-63.14$ & $-62.84$ & $-75.51$ & $-75.04$ \\
   & $1d$ & $-21.90$ & $-22.10$ & $-31.57$ & $-31.34$ & $-45.03$ & $-44.36$ & $-58.98$ & $-57.96$ & $-71.26$ & $-69.98$ \\
   & $2s$ & $-21.37$ & $-21.67$ & $-30.71$ & $-30.60$ & $-43.88$ & $-43.40$ & $-57.58$ & $-56.85$ & $-69.67$ & $-68.77$ \\
  \midrule
  ${}^{208}\mathrm{Pb}$ & $1s$ & $-29.87$ & $-29.98$ & $-40.26$ & $-40.00$ & $-54.41$ & $-53.70$ & $-68.97$ & $-67.79$ & $-81.74$ & $-80.14$ \\
   & $1p$ & $-26.40$ & $-26.52$ & $-36.54$ & $-36.27$ & $-50.48$ & $-49.71$ & $-64.85$ & $-63.61$ & $-77.48$ & $-75.83$ \\
   & $1d$ & $-22.74$ & $-22.83$ & $-32.57$ & $-32.14$ & $-46.23$ & $-45.22$ & $-60.38$ & $-58.86$ & $-72.85$ & $-70.91$ \\
   & $2s$ & $-22.46$ & $-22.41$ & $-32.12$ & $-31.43$ & $-45.69$ & $-44.30$ & $-59.79$ & $-57.80$ & $-72.24$ & $-69.75$ \\
  \bottomrule\bottomrule
 \end{tabular*}
\end{table*}

\begin{table*}[!t]
 \caption{Charged $\Occc$ energies. The reference entries underlie Fig.~\ref{fig:omega-omegaccc-level-spectra}(b). The cosh calculation uses the same geometry as for $\Omega$, with the $\Occc$ mass, nuclear-matter depths, and charge $+2$. All energies are in MeV and are rounded to two decimal places. Each cutoff has paired reference (Ref.) and cosh columns. A dash denotes no bound state in the specified orbital; $\dagger$ marks a difference in bound-state classification.}
 \label{tab:app-ccc-charged}
 \centering
 \setlength{\tabcolsep}{3pt}
 \begin{tabular*}{\textwidth}{@{\extracolsep{\fill}}cc*{10}{c}}
  \toprule\toprule
  Core & State & \multicolumn{2}{c}{$\alpha=1$} & \multicolumn{2}{c}{$\alpha=1.5$} & \multicolumn{2}{c}{$\alpha=2$} & \multicolumn{2}{c}{$\alpha=2.5$} & \multicolumn{2}{c}{$\alpha=3$} \\
  \cmidrule(lr){3-4} \cmidrule(lr){5-6} \cmidrule(lr){7-8} \cmidrule(lr){9-10} \cmidrule(lr){11-12}
  & & Ref. & Cosh & Ref. & Cosh & Ref. & Cosh & Ref. & Cosh & Ref. & Cosh \\
  \midrule
  ${}^{12}\mathrm{C}$ & $1s$ & \textemdash & \textemdash & $-3.17$ & $-4.59$ & $-17.05$ & $-18.79$ & $-30.61$ & $-32.44$ & $-41.46$ & $-43.31$ \\
   & $1p$ & \textemdash & \textemdash & \textemdash & \textemdash & $-8.78$ & $-10.51$ & $-20.60$ & $-22.60$ & $-30.36$ & $-32.48$ \\
   & $1d$ & \textemdash & \textemdash & \textemdash & \textemdash & $-0.83$ & $-2.15$ & $-10.55$ & $-12.37$ & $-19.05$ & $-21.11$ \\
   & $2s$ & \textemdash & \textemdash & \textemdash & \textemdash & $-1.31$ & $-2.08$ & $-10.52$ & $-11.77$ & $-18.76$ & $-20.25$ \\
  \midrule
  ${}^{16}\mathrm{O}$ & $1s$ & \textemdash & \textemdash & $-3.30$ & $-4.05$ & $-18.32$ & $-18.84$ & $-32.66$ & $-32.88$ & $-44.02$ & $-44.03$ \\
   & $1p$ & \textemdash & \textemdash & \textemdash & \textemdash & $-11.58$ & $-11.75$ & $-24.81$ & $-24.56$ & $-35.51$ & $-34.92$ \\
   & $1d$ & \textemdash & \textemdash & \textemdash & \textemdash & $-4.34$ & $-4.29$ & $-16.14$ & $-15.65$ & $-25.99$ & $-25.12$ \\
   & $2s$ & \textemdash & \textemdash & \textemdash & \textemdash & $-3.55$ & $-3.71$ & $-14.69$ & $-14.67$ & $-24.16$ & $-23.93$ \\
  \midrule
  ${}^{40}\mathrm{Ca}$ & $1s$ & \textemdash & \textemdash & \textemdash & \textemdash & $-15.39$ & $-13.71$ & $-31.35$ & $-28.55$ & $-43.94$ & $-40.22$ \\
   & $1p$ & \textemdash & \textemdash & \textemdash & \textemdash & $-11.50$ & $-9.88$ & $-26.67$ & $-24.13$ & $-38.71$ & $-35.43$ \\
   & $1d$ & \textemdash & \textemdash & \textemdash & \textemdash & $-7.25$ & $-5.42$ & $-21.69$ & $-18.99$ & $-33.26$ & $-29.88$ \\
   & $2s$ & \textemdash & \textemdash & \textemdash & \textemdash & $-6.59$ & $-4.15$ & $-21.20$ & $-17.56$ & $-32.93$ & $-28.36$ \\
  \midrule
  ${}^{90}\mathrm{Zr}$ & $1s$ & \textemdash & \textemdash & \textemdash & \textemdash & $-6.17$ & $-4.40$ & $-21.97$ & $-19.48$ & $-34.42$ & $-31.33$ \\
   & $1p$ & \textemdash & \textemdash & \textemdash & \textemdash & $-4.61$ & $-2.60$ & $-20.04$ & $-17.37$ & $-32.25$ & $-29.04$ \\
   & $1d$ & \textemdash & \textemdash & \textemdash & \textemdash & $-2.52$ & $-0.20$ & $-17.57$ & $-14.63$ & $-29.53$ & $-26.09$ \\
   & $2s$ & \textemdash & \textemdash & \textemdash & \textemdash & $-1.11$ & \textemdash$^{\dagger}$ & $-16.25$ & $-13.03$ & $-28.24$ & $-24.48$ \\
  \midrule
  ${}^{197}\mathrm{Au}$ & $1s$ & \textemdash & \textemdash & \textemdash & \textemdash & \textemdash & \textemdash & $-6.63$ & $-4.01$ & $-18.30$ & $-15.80$ \\
   & $1p$ & \textemdash & \textemdash & \textemdash & \textemdash & \textemdash & \textemdash & $-6.17$ & $-3.29$ & $-17.81$ & $-15.01$ \\
   & $1d$ & \textemdash & \textemdash & \textemdash & \textemdash & \textemdash & \textemdash & $-5.31$ & $-2.07$ & $-16.90$ & $-13.70$ \\
   & $2s$ & \textemdash & \textemdash & \textemdash & \textemdash & \textemdash & \textemdash & $-1.84$ & \textemdash$^{\dagger}$ & $-13.47$ & $-11.27$ \\
  \midrule
  ${}^{208}\mathrm{Pb}$ & $1s$ & \textemdash & \textemdash & \textemdash & \textemdash & \textemdash & \textemdash & $-5.87$ & $-3.16$ & $-17.72$ & $-14.94$ \\
   & $1p$ & \textemdash & \textemdash & \textemdash & \textemdash & \textemdash & \textemdash & $-5.36$ & $-2.49$ & $-17.14$ & $-14.21$ \\
   & $1d$ & \textemdash & \textemdash & \textemdash & \textemdash & \textemdash & \textemdash & $-4.46$ & $-1.35$ & $-16.16$ & $-12.98$ \\
   & $2s$ & \textemdash & \textemdash & \textemdash & \textemdash & \textemdash & \textemdash & $-1.86$ & \textemdash$^{\dagger}$ & $-13.81$ & $-10.45$ \\
  \bottomrule\bottomrule
 \end{tabular*}
\end{table*}

\interlinepenalty=10000
\bibliography{references}
\end{document}